\documentclass[a4paper, 10pt]{report}

\usepackage[cp1250]{inputenc}
\usepackage[T1]{fontenc}
\usepackage{amsfonts}
\usepackage{graphicx}
\usepackage{amsmath}
\usepackage{caption}

\usepackage{amssymb}

\usepackage{fancyhdr}
\usepackage{bibtopic}
\usepackage{color}
\usepackage{ulem}
\usepackage[toc,page]{appendix}
\usepackage{setspace}
\usepackage{cite}
\usepackage{tensor}

\AtBeginDocument{}
\usepackage{etoolbox}
\patchcmd{\thebibliography}{\chapter*}{\section*}{}{}

\makeatletter
\renewcommand{\thesection}{%
  \ifnum\c@chapter<1 \@arabic\c@section
  \else \thechapter.\@arabic\c@section
  \fi
}
\makeatother

\patchcmd{\tableofcontents}{\chapter*}{\section*}{}{}

\numberwithin{equation}{section}

\let\OLDthebibliography\thebibliography
\renewcommand\thebibliography[1]{
  \OLDthebibliography{#1}
  \setlength{\parskip}{0pt}
  \setlength{\itemsep}{3.5pt plus 1ex}
}

\usepackage[linktocpage]{hyperref}
\definecolor{darkred}{rgb}{0.5,0,0}
\definecolor{darkpurple}{rgb}{0.5,0,0.5}
\definecolor{darkblue}{rgb}{0,0,0.5}
\hypersetup{ colorlinks,
linkcolor=darkblue,
filecolor=darkpurple,
urlcolor=darkred,
citecolor=darkpurple }
 
\usepackage[symbol]{footmisc}
\usepackage{cancel}

\begin{document}

\allowdisplaybreaks
\setlength{\abovedisplayskip}{3.5pt}
\setlength{\belowdisplayskip}{3.5pt}
\abovedisplayshortskip
\belowdisplayshortskip

{\setstretch{1.0}

{\LARGE \bf \centerline{
Evolution of superhorizon perturbations
}}
{\LARGE \bf \centerline{
in early Universe with anisotropic solid remnant
}}

\vskip 1cm
\begin{center}
{Peter M\'esz\'aros\footnote[1]{peter.meszaros@fmph.uniba.sk}, Daniel Ra\v{c}ko\footnote[2]{daniel.racko@fmph.uniba.sk}}

\vskip 2mm {\it Department of Theoretical Physics, Comenius
University, Bratislava, Slovakia}

\vskip 2mm \today 
\end{center}

\section*{Abstract}
We study effects of presence of a small amount of so-called anisotropic solid remnant in the early post-inflationary Universe dominated by radiation.
This model is inspired by solid inflation and its generalizations with matter described through a triplet of fields.
In our model, the internal full global Euclidean symmetry of this triplet is broken, which leads to an anisotropic expansion of the Universe.
Superhorizon scalar and vector perturbations grow, while behavior of tensor perturbations remains the same as in the standard case with Universe filled with only radiation.
We also find a very interesting case within the limit of a very small amount of the anisotropic solid remnant and the longest possible duration of its presence, where size of vector perturbations decreases with a different power of scale factor as in the standard scenario.
The obtained results improve agreement between the observational data and the theoretical predictions of solid inflation in a case with suppressed non-linear effects.

\vskip 3mm \hspace{4mm}
\begin{minipage}[t]{0.8\textwidth}
\noindent\rule{12cm}{0.4pt}
\vspace{-9mm}
\tableofcontents
\noindent\rule{12cm}{0.4pt}
\end{minipage}
\vskip 6mm

}

\section{Introduction}\label{sec:1}

$\Lambda$CDM cosmology, together with models of cosmic inflation\cite{guth,linde} and reheating\cite{kofman1,shtanov,kofman2}, provides a very successful framework for determining the properties of the Universe from observational data\cite{planck1,planck2}. However, there are still some anomalies\cite{schwarz}. The most important problem is the tension between the local value of the Hubble constant and its value from the cosmic microwave background (CMB)\cite{riess1,riess2,divalentino,freedman}, and there are also anomalies of the CMB anisotropies for low multipole moments\cite{land,pinkwart}. Considering an anisotropic expansion of the Universe may address these problems. Anisotropy with Bianchi-type backgrounds may be related to such anomalies\cite{pontzen}, and other types of background anisotropy \cite{chang} may be caused by cosmic strings \cite{yang,jazayeri} or domain walls \cite{jazayeri2,firouzjahi}. Another approach to this problem is based on cosmological models with bounce \cite{agullo}.

\vskip 2mm
Theory of cosmological perturbations is the key tool which enables comparison between theoretical models and observations of CMB anisotropies. Of three kinds of perturbations, scalar, vector, and tensor perturbations\cite{bardeen}, only the scalar sector appears to have the dominant contribution\cite{cmbslow}. Tensor perturbations are negligible due to the small tensor-to-scalar ratio\cite{planck2,bicep}, and vector perturbations are strongly suppressed in the course of expansion of the Universe. Even in some models where vector fields drive the inflation, vector perturbations remain negligible\cite{golovnev}. Vector perturbations do not have a strong effect on CMB anisotropies\cite{saga} and on the distribution of galaxies\cite{durrer}, however, they may be responsible for the aforementioned CMB anomalies\cite{morales,chen}. Since vector perturbations are being suppressed as the Universe expands, the opposite is also true, and in models with bounce one has to treat them with caution\cite{battefeld,zhu}.

\vskip 2mm
In this work, we focus on the case with Bianchi type-I background anisotropy. This may be the case when inflation driven by inflaton fields with global Euclidean symmetry $E(3)=SO(3)\ltimes T(3)$, so-called solid inflation\cite{gruzinov,endlich}, ends by a phase transition in which most of the inflaton energy transfers to radiation described as a perfect fluid, while a small portion of it remains in the form of the original inflaton fields, however, with a broken symmetry $E(3)\to SO(2)\ltimes T(3)$. We will call this anisotropic matter component which originates from the inflaton fields the anisotropic solid remnant. This represents a generalization of the reheating mechanism proposed in\cite{endlich}. Based on this motivation, we will study a simplified scenario in which there is a short period when the Universe is filled with only two matter components, the radiation fluid and the anisotropic solid remnant.

\vskip 2mm
To analyze the evolution of cosmological perturbations, the standard approach with the isotropic background\cite{bardeen} has to be replaced by parameterization of perturbations on the anisotropic background\cite{gumru}.
Such parameterization allows for existence of only two types of perturbations, scalar perturbations transforming as scalars, and vector perturbations that transform as two-dimensional vectors under the residual $SO(2) \ltimes T(3)$ symmetry. In the limit of vanishing background anisotropy, one can retrieve the standard three-dimensional scalar, vector and tensor perturbations, however, they are coupled to each other in contrast to the standard case with the isotropic background. Such violation of the decomposition theorem has been found out also through a different formalism\cite{anizpicon,pereira} in which scalar, vector and tensor perturbations are defined directly on the general Bianchi type-I background. Since we are studying the case with the residual $SO(2) \ltimes T(3)$ symmetry in this work, we can use the former formalism\cite{gumru} with decoupled two-dimensional scalar and vector sectors.
We will treat perturbations classically, since the observationally relevant modes were superhorizon after the end of inflation, and we will focus on the superhorizon behavior of modes. We believe, that analysis of the model proposed in this paper enables us to capture the main features of evolution of perturbations and effect of the background anisotropy on it. Because of the background anisotropy, the vector perturbations deserve a special attention.

\vskip 2mm
In the next section \ref{sec:2}, we formulate the model under consideration and describe its relation to the solid inflation model. Sections \ref{sec:3} and \ref{sec:4} will be dedicated to the background expansion and evolution of superhorizon perturbations, respectively. In section \ref{sec:5}, we analyze theoretical predictions of the studied model from the point of view of cosmological observations, and in the last section \ref{sec:6} we summarize the results. Throughout the work, we use units in which $c=1$ while denoting the Newtonian gravitational constant as $\kappa$, and the signature of the spacetime metric is chosen as $-,+,+,+$. Four spacetime indices will be denoted by Greek letters, space indices by small Latin letters, and capital Latin indices will run through only two space coordinates associated with the remaining $SO(2)$ symmetry, $\mu=0,i$, $i=1,A$, $A=2,3$.

\section{Model under consideration}\label{sec:2}

We assume that the early Universe dominated by radiation temporarily contained also an exotic matter component with anisotropic properties. Such anisotropic matter component may be a remnant of a triplet of inflaton fields $\Phi^i$ driving the solid inflation\cite{endlich}. The matter Lagrangian of this model is
\begin{eqnarray}\label{eq:solidinflation}
\mathcal{L}_{\textrm{m}}=F(X,Y,Z), \quad X=\textrm{Tr}B, \quad Y=\dfrac{\textrm{Tr}B^2}{X^2}, \quad Z=\dfrac{\textrm{Tr}B^3}{X^3}, \quad B^{ij}=g^{\mu\nu}\Phi^i_{\phantom{i},\mu}\Phi^j_{\phantom{j},\nu},
\end{eqnarray}
the background configuration of fields is $\Phi^i=c x^i$, where $x^i$ are comoving Cartesian space coordinates, and the background spacetime configuration is given by the Friedmann--Lema\^itre--Robertson--Walker (FLRW) metric with flat space geometry,
\begin{eqnarray}\label{eq:flrw}
ds^2 = -dt^2 + a(t)^2 \delta_{ij} dx^i dx^j = a(\tau)^2\left(-d\tau^2+\delta_{ij}dx^i dx^j\right),
\end{eqnarray}
where either the cosmic time $t$ or conformal time $\tau$ may be used as a time coordinate, $x^i$ are aforementioned space coordinates, and $a$ is the scale factor. Quantities $X$, $Y$ and $Z$ are invariant under global translations and rotations, and therefore, the symmetry of such model is the global Euclidean symmetry $E(3)=SO(3)\ltimes T(3)$.

\vskip 2mm
Authors of\cite{endlich} proposed a reheating mechanism based on the following form of the matter Lagrangian
\begin{eqnarray}\label{eq:twosegments}
\mathcal{L}_{\textrm{m}}=F(X,Y,Z)\propto\left\{\begin{array}{lll}
(\textrm{det}B)^{\epsilon/3}f(Y,Z) & \textrm{for} & a < a_{\textrm{inf.}} \\
(\textrm{det}B)^{2/3} & \textrm{for} & a > a_{\textrm{inf.}}
\end{array}\right.,
\end{eqnarray}
where $\epsilon$ is the Hubble-flow parameter during inflation defined as $\epsilon=-\dot{H}/H^2$, $H=\dot{a}/a$, with dot denoting differentiation with respect to cosmic time $t$, and the process of reheating is controlled by the value of $\textrm{det}B = c^6 a^{-6}$. When $\textrm{det}B$ drops bellow $c^6 a_{\textrm{inf.}}^{-6}$, with $a_{\textrm{inf.}}$ being the scale factor at the end of inflation, the solid inflation transits to the reheating era with matter Lagrangian indistinguishable from the case with radiation-type perfect fluid. Both phases have the same physical degrees of freedom, but different internal symmetries. The Euclidean internal symmetry valid during solid inflation changes to a larger internal symmetry of all volume-preserving diffeormorphisms.

\vskip 2mm The solid inflation model had been studied also in the case with anisotropic background\cite{bartolo1,bartolo2}, corresponding to a special case of Bianchi type-I spacetime metric,
\begin{eqnarray}\label{eq:bianchi}
ds^2 = -dt^2 + a(t)^2 (dx^1)^2 + b(t)^2 \delta_{AB} dx^A dx^B,
\end{eqnarray}
where $a(t)$ and $b(t)$ describe two different direction-dependent rates of the expansion, with residual isotropy in the $\{x^2,x^3\}$-plane. The Bianchi type-I metric can also be used to describe a more general case with three different rates of the expansion with no residual isotropy, but we will not study this case in this work. Authors of\cite{bartolo1} found violation of cosmic no-hair conjecture\cite{wald} in the solid inflation model, however, the anisotropy is still decreasing in the course of the expansion.

\vskip 2mm
We are interested in introducing anisotropy not only to the spacetime metric, but also to the form of the matter Lagrangian. The simplest way is to replace the quantity $X$ invariant with respect to global Euclidean symmetry with two quantities $\Theta$ and $\Omega$ associated with a smaller $SO(2)\ltimes T(3)$ symmetry, defined as
\begin{eqnarray}\label{eq:thetaomega}
\Theta = g^{\mu\nu} \Phi^1_{\phantom{1},\mu} \Phi^1_{\phantom{1},\nu}, \quad
\Omega = \dfrac{1}{2} g^{\mu\nu} \delta_{AB} \Phi^A_{\phantom{A},\mu} \Phi^B_{\phantom{B},\nu}.
\end{eqnarray}
We can see that $X=\Theta+2\Omega$, which means that a matter Lagrangian that depends only on this combination of $\Theta$ and $\Omega$ is fully symmetric with respect to global Euclidean symmetry as long as the background spacetime metric is the FLRW metric (\ref{eq:flrw}). However, in the case with the Bianchi type-I metric of the form (\ref{eq:bianchi}), we have $\Theta=c^2 a^{-2}$ and $\Omega=c^2 b^{-2}$ for the background configuration. Let us now assume the following transition from inflationary expansion to the beginning of the standard $\Lambda$CDM phase dominated by radiation in the form of a perfect fluid,
\begin{eqnarray}\label{eq:segments}
\mathcal{L}_{\textrm{m}}=F(X,Y,Z)\propto\left\{\begin{array}{lll}
X^{\epsilon} = \left(\Theta+2\Omega\right)^{\epsilon} & \textrm{for} & a < a_{\textrm{inf.}} \\
\mathcal{L}_{\textrm{rad.}} + \lambda f(\Theta,\Omega) & \textrm{for} & a_{\textrm{inf.}} < a < a_{\textrm{reh.}} \\
\mathcal{L}_{\textrm{rad.}} & \textrm{for} & a_{\textrm{reh.}} < a
\end{array}\right.,
\end{eqnarray}
with $a_{\textrm{reh.}}$ denoting the scale factor at which the reheating period ends and the standard $\Lambda$CDM period begins. Parameter $\epsilon$ is again the Hubble-flow parameter, $\lambda$ is a constant which qualitatively measures the amount of the solid remnant admixture, and $\mathcal{L}_{\textrm{rad.}}$ is matter Lagrangian of the radiation component. Here we have three different time periods:
\begin{itemize}
\item During the inflationary period ($a < a_{\textrm{inf.}}$), the radiation in the form of perfect fluid is either not present, or the fluid component started to gain significant amount of energy by transfer from the inflaton fields only after all observationally relevant modes of perturbations were already superhorizon and they were, up to a small rate of change given by slow-roll parameters, conserved in the superhorizon limit. This assumption preserves validity of predictions of the solid inflation model based on analysis with perturbations treated in the framework of quantum field theory on the curved background\cite{endlich}. The process of energy transfer from inflaton fields to radiation is studied within warm inflation\cite{warm1,warm2,warm3}. When $4$-velocities associated with inflaton and radiation components differ from each other, the expansion is anisotropic\cite{warmaniz}.
\item After the end of inflation ($a_{\textrm{inf.}} < a < a_{\textrm{reh.}}$), the radiation is the dominant component, while inflaton fields remain in the form of solid remnant with anisotropic properties. Dominance of radiation component can be achieved by the choice of constant $\lambda$ in (\ref{eq:segments}) such that it is small, $\lambda\ll 1$. Such anisotropy may be caused by anisotropy during the solid inflation era, which should be small during inflation to be consistent with observations\cite{bartolo1,bartolo2}. Since all observationally relevant modes became superhorizon during the inflationary time period, perturbations may be treated clasically. If this period is short enough, the superhorizon modes do not have enough time to reenter the horizon, and it is sufficient to study perturbations only in the superhorizon limit.
\item After the end of reheating ($a_{\textrm{reh.}} < a$), the standard $\Lambda$CDM model serves as a full description of the evolution of the background expansion as well as cosmological perturbations leaving an imprint on the CMB anisotropies. Models of the early Universe during this era with Bianchi type-I background have been analyzed in light of observational data in several works\cite{kim,schucker,akarsu,verma,herzberg}. A transition from the previous period may be due to another phase transition when the energy level drops in the course of the expansion, or a result of non-linear effects enhanced by growth of perturbations, which will be discussed in sections \ref{sec:4} and \ref{sec:5}.
\end{itemize}
The key difference between the reheating process given by (\ref{eq:twosegments}) and the newly proposed scenario (\ref{eq:segments}) is that in the first case the triplet of inflaton fields carry all degrees of freedom of the system, while in the second case the number of degrees of freedom is larger due to the presence of the radiation fluid. Of course, the reheating scenario considered above represents a very simplified version of a more realistic transition process from inflation to the standard $\Lambda$CDM model dominated by radiation in early times. However, we believe that by studying a cosmological model with two matter components, $\mathcal{L}_{\textrm{m}}=\mathcal{L}_{\textrm{rad.}}+\lambda f(\Theta,\Omega)$, we capture the qualitative nature of post-inflationary perturbations and the role of the background anisotropy in this process.

\vskip 2mm
The stress-energy tensor of the system under consideration consists of two parts. For the radiation part, we have the standard relation
\begin{eqnarray}
T^{(\textrm{rad.})}_{\mu\nu} = \rho \left( \dfrac{4}{3} u_{\mu}u_{\nu} + \dfrac{1}{3} g_{\mu\nu} \right),
\end{eqnarray}
where $\rho$ is the energy density of the radiation component and $u_{\mu}$ is $4$-velocity of its volume elements. The part corresponding to fields $\Phi^i$ can be calculated from the canonical stress-energy tensor as
\begin{eqnarray}
T^{(\Phi)}_{\mu\nu} = - 2\dfrac{\delta f}{\delta g^{\mu\nu}} + f g_{\mu\nu} = -2f_{\Theta}\Phi^1_{\phantom{1},\mu}\Phi^1_{\phantom{1},\nu} - f_{\Omega}\delta_{AB}\Phi^A_{\phantom{A},\mu}\Phi^B_{\phantom{B},\mu} + f g_{\mu\nu},
\end{eqnarray}
where $f_{\Theta}$ and $f_{\Omega}$ denote partial derivatives of function $f(\Theta,\Omega)$ with respect to $\Theta$ and $\Omega$. The overall stress-energy tensor is then
\begin{eqnarray}
T_{\mu\nu} = T^{(\textrm{rad.})}_{\mu\nu} + \lambda T^{(\Phi)}_{\mu\nu},
\end{eqnarray}
with constant $\lambda$ controlling admixture of the anisotropic solid remnant. We will assume a smallness of the parameter $\lambda$ such that we can control the perturbative expansion of quantities through it. Our calculations will be only up to the first order in parameter $\lambda$, and for this purpose we will use the notation $Q = Q_0 + \lambda Q_1 + \mathcal{O}(\lambda^2)$ for any quantity $Q$ under consideration. Therefore, for the stress-energy tensor we have
\begin{eqnarray}
(T_0)_{\mu\nu} = (T_0)^{(\textrm{rad.})}_{\mu\nu}, \quad
(T_1)_{\mu\nu} = (T_1)^{(\textrm{rad.})}_{\mu\nu} + (T_0)^{(\Phi)}_{\mu\nu},
\end{eqnarray}
where the relation for $(T_0)_{\mu\nu}$ simply means that the solid remnant has no effect in the zeroth order in $\lambda$. The part $(T_1)^{(\textrm{rad.})}_{\mu\nu}$ arises due to the decomposition of the radiation energy density and $4$-velocity, $\rho = \rho_0+\lambda\rho_1$, $u_{\mu}=(u_0)_{\mu}+\lambda (u_1)_{\mu}$. All zero-order quantities $Q_0$ should be the same as in the radiation-dominated era, while the first-order parts $Q_1$ will measure the leading order effect of the presence of the anisotropic solid remnant.

\vskip 2mm
This perturbative expansion is independent from the expansion associated with cosmological perturbations. Both the background quantities, for example scale factor, as well as cosmological perturbations, for example tensor perturbations, will be subjects to the perturbative decomposition with respect to the parameter $\lambda$.

\section{Background solutions}\label{sec:3}

In order to examine evolution of the unperturbed Universe and determine the anisotropic effect of the solid remnant, it is sufficient to use only Einstein field equations. It is also convenient to parameterize the unperturbed Bianchi type-I spacetime (\ref{eq:bianchi}) as
\begin{eqnarray}\label{eq:background}
ds^2 = e^{2\alpha} \left(-d\tau^2 + e^{-4\sigma} (dx^1)^2 + e^{2\sigma} \delta_{AB}dx^Adx^B\right),
\end{eqnarray}
where $\alpha$ and $\sigma$ depend on conformal time $\tau$. Physical volumes of comoving regions are proportional to $e^{3\alpha}$, and therefore, it is reasonable to call $e^{\alpha}$ the scale factor, $a=e^{\alpha}$, while the function $\sigma$ measures the size of the anisotropy through a conventionally defined quantity $\gamma=\dot{\sigma}/\dot{\alpha}$. In order to have an anisotropic spacetime different from the isotropic FLRW spacetime only rewritten through some rescaled coordinates, the parameter $\gamma$ cannot be zero. If the anisotropy is caused by only the anisotropic remnant, we can set $\sigma = \lambda \sigma_1$, which means that without its presence we retrieve the isotropic FLRW background with $\sigma = \sigma_0 = 0$.

\vskip 2mm
All nonzero components of the Einstein tensor corresponding to the spacetime metric (\ref{eq:background}) in the zeroth order in $\lambda$ are
\begin{eqnarray}\label{eq:G0}
(G_0)_{00} = 3\alpha_0^{\prime 2}, \quad (G_0)_{ij} = - \left( \alpha_0^{\prime 2} + 2 \alpha_0^{\prime\prime}\right) \delta_{ij},
\end{eqnarray}
where $0$ components are associated with the conformal time $\tau$ and prime denotes differentiation with respect to it. Components in the first order in $\lambda$ read
\begin{eqnarray}\label{eq:G1}
& & (G_1)_{00} = 6 \alpha_0^{\prime} \alpha_1^{\prime}, \\
& & (G_1)_{11} = 4 \left(\alpha_0^{\prime 2} + 2\alpha_0^{\prime\prime}\right) \sigma_1 - 2 \alpha_0^{\prime} \alpha_1^{\prime} - 4 \alpha_0^{\prime}\sigma_1^{\prime} - 2 \alpha_1^{\prime\prime} - 2 \sigma_1^{\prime\prime}, \nonumber\\
& & (G_1)_{AB} = \left[ -2 \left(\alpha_0^{\prime 2} + 2\alpha_0^{\prime\prime}\right) \sigma_1 - 2 \alpha_0^{\prime} \alpha_1^{\prime} + 2 \alpha_0^{\prime}\sigma_1^{\prime} - 2 \alpha_1^{\prime\prime} + \sigma_1^{\prime\prime} \right] \delta_{AB}. \nonumber
\end{eqnarray}
For the stress-energy tensor entering the right-hand side of Einstein field equations we have
\begin{eqnarray}\label{eq:T0}
(T_0)_{00} = e^{2\alpha_0} \rho_0, \quad (T_0)_{ij} = \frac{1}{3} e^{2\alpha_0} \rho_0 \delta_{ij},
\end{eqnarray}
and
\begin{eqnarray}\label{eq:T1}
& & (T_1)_{00} = e^{2\alpha_0} \rho_0 \left[ (2\alpha_1+\delta_1) +q_I\right], \\
& & (T_1)_{11} = e^{2\alpha_0} \rho_0 \left[ \frac{1}{3}(2\alpha_1+\delta_1) -\frac{4}{3} \sigma_1 -q_I + q_1 \right], \nonumber\\
& & (T_1)_{AB} = e^{2\alpha_0} \rho_0 \left[ \frac{1}{3}(2\alpha_1+\delta_1) +\frac{2}{3} \sigma_1 -q_I + q_{\perp} \right]\delta_{AB}, \nonumber
\end{eqnarray}
where $\delta_1 = \rho_1/\rho_0$ is defined by ratio of quantities describing radiation energy density,
\begin{eqnarray}
\rho = \rho_0 + \lambda \rho_1 = \rho_0 \left( 1 + \lambda \delta_1 \right),
\end{eqnarray}
and coefficients $q_I$, $q_1$ and $q_{\perp}$ are given by partial derivatives of $f(\Theta,\Omega)$ as
\begin{eqnarray}
q_I = \dfrac{f^0}{\rho_0}, \quad q_1 = 2 \dfrac{\Theta_0 f^0_{\Theta}}{\rho_0}, \quad q_{\perp} = \dfrac{\Omega_0 f^0_{\Omega}}{\rho_0},
\end{eqnarray}
with the upper index $0$ at $f^0$, $f^0_{\Theta}$ and $f^0_{\Omega}$ standing for evaluation of function $f$ and its derivatives at $\Theta_0=\Omega_0=c^2 e^{-2\alpha_0}$ corresponding to the zeroth order in the expansion with respect to parameter $\lambda$.

\vskip 2mm
Einstein field equations in the zeroth order in $\lambda$, $(G_0)_{\mu\nu}=8\pi\kappa(T_0)_{\mu\nu}$, can be derived from (\ref{eq:G0}) and (\ref{eq:T0}). This yields two independent equations for quantities $\alpha_0$ and $\rho_0$, the Friedmann equation and the acceleration equation,
\begin{eqnarray}
3\alpha_0^{\prime 2} = 8\pi\kappa e^{2\alpha_0}\rho_0, \quad -\alpha_0^{\prime 2} - 2\alpha_0^{\prime\prime} = \frac{8\pi\kappa}{3} e^{2\alpha_0}\rho_0.
\end{eqnarray}
Their solution is
\begin{eqnarray}\label{eq:sol0}
\alpha_0 = \alpha_{*} + \ln \dfrac{\tau}{\tau_{*}}, \quad \rho_0 = \dfrac{3 e^{-2\alpha_{*}} \tau_{*}^2}{8\pi\kappa} \tau^{-4},
\end{eqnarray}
where $\tau_{*}$ is a reference conformal time at which the value of the scale factor is $a_{*}=e^{\alpha_{*}}$. Since the anisotropic solid remnant does not affect zeroth-order quantities $Q_0$, this is the usual result corresponding to an isotropic and flat Universe with radiation, which is also in agreement with the fluid equation for radiation, $T_{\mu\nu}^{\phantom{\mu\nu};\mu}u^{\nu} = 0$ $\Rightarrow$ $\rho_0^{\prime} + 4 \alpha_0^{\prime}\rho_0 = 0$, and the usual relation $\rho_0 \propto a_0^{-4}$.

\vskip 2mm
In the first order in $\lambda$, we have three independent equations for $\alpha_1$, $\sigma_1$ and $\delta_1$ which follow from comparing three independent components of the Einstein tensor (\ref{eq:G1}) with components of the stress-energy tensor (\ref{eq:T1}), $(G_1)_{\mu\nu}=8\pi\kappa(T_1)_{\mu\nu}$. It is easy to find decoupled versions of these equations by using appropriate combinations of Einstein field equations,
\begin{eqnarray}
\label{eq:alpha1}
& & (G_1)_{00} - (G_1)_{ii} = 8\pi\kappa \left( (T_1)_{00} - (T_1)_{ii} \right) \Rightarrow \nonumber\\
& & \alpha_1^{\prime\prime} + \dfrac{2}{\tau} \alpha_1^{\prime} = \dfrac{8\pi\kappa}{3} c^2 \left( 2\dfrac{f^0}{\sqrt{\Theta_0\Omega_0}} - f^0_{\Theta} - f^0_{\Omega} \right), \\
\label{eq:sigma1}
& & 2(G_1)_{11} - (G_1)_{22} - (G_1)_{33} = 8\pi\kappa \left( 2(T_1)_{11} - (T_1)_{22} - (T_1)_{33} \right) \Rightarrow \nonumber\\
& & \sigma_1^{\prime\prime} + \dfrac{2}{\tau} \sigma_1^{\prime} = - \dfrac{8\pi\kappa}{3} c^2 \left( 2f^0_{\Theta} - f^0_{\Omega} \right), \\
\label{eq:delta1}
& & (G_1)_{00} = 8\pi\kappa (T_1)_{00} \Rightarrow \nonumber\\
& & \delta_1 = - 2\alpha_1 + 2\tau\alpha_1^{\prime} - \dfrac{8\pi\kappa}{3} c^2 \tau^2 \dfrac{f^0}{\sqrt{\Theta_0\Omega_0}},
\end{eqnarray}
where the solution from the previous order (\ref{eq:sol0}) has been used. All right-hand sides of these equations are functions of conformal time, where $\Theta_0=\Omega_0=c^2 a_{*}^{-2} (\tau/\tau_{*})^{-2}$, which, of course, depend on the form of function $f$.

\vskip 2mm
In order to achieve conclusions when analyzing the model under consideration, we will consider the following form of function $f$ determining the matter Lagrangian of the anisotropic solid remnant,
\begin{eqnarray}\label{eq:PQ}
f(\Theta,\Omega) = K \Theta^P \Omega^Q,
\end{eqnarray}
where $K$, $P$ and $Q$ are constants. Since equations (\ref{eq:alpha1})-(\ref{eq:delta1}) are linear, their solution contains homogeneous and particular parts. Because of the linearity of their right-hand sides with respect to the function $f$, one can consider the function $f$ to be an arbitrary sum of functions of the form (\ref{eq:PQ}), which covers a wide range of possible functions, and then one can use the particular solution corresponding to (\ref{eq:PQ}) to write down the particular solution for a more general form of the function $f$. Therefore, by restricting ourselves to the case with (\ref{eq:PQ}) we not only considerably simplify following calculations, but we also keep a reasonable level of generality.

\vskip 2mm
Left-hand-sides of equations (\ref{eq:alpha1})-(\ref{eq:delta1}) can be rewritten into a more convenient form by substituting the conformal time $\tau$ with the logarithm of the scale factor. Here we choose
\begin{eqnarray}\label{eq:defy}
e^y = \dfrac{a_0}{a} = q \dfrac{a_0 H_0}{\tau},
\end{eqnarray}
where $a_0$ and $H_0$ are the scale factor and Hubble parameter today, the second equality holds during the radiation dominated era, and $q$ is a number that depends on details of the expansion up to the present time. At the present time, we have $y=0$, and $y$ is decreasing with increasing time. This enables us to connect the conformal time with the energy scale associated with the radiation component of the matter. By rewriting (\ref{eq:defy}) in terms of temperature, $e^y = T_0/T$, where $T_0 = 2.726 \textrm{T}$, we find that at the Grand Unified Theory (GUT) energy scale, $ k_{\textrm{B}}T=10^{25}\textrm{eV}$, we have $y=65.92$, and for example at the $100 \textrm{TeV}$ energy scale, which is beyond the reach of current particle accelerators, we have $y=40.59$. If the inflation occurs at GUT energy scale, this is the most reasonable range during which the model studied in this paper may be relevant.

\vskip 2mm
By considering the function $f$ to be of the form (\ref{eq:PQ}) and using variable $y$ defined by (\ref{eq:defy}), equations (\ref{eq:alpha1})-(\ref{eq:delta1}) can be rewritten as
\begin{eqnarray}
\label{eq:alpha11}
& & \dfrac{d^2 \alpha_1}{dy^2} - \dfrac{d\alpha_1}{dy} = \tilde{\mathcal{C}} (2-P-Q) e^{-2(2-P-Q)y}, \\
\label{eq:sigma11}
& & \dfrac{d^2 \sigma_1}{dy^2} - \dfrac{d\sigma_1}{dy} = \tilde{\mathcal{C}} (Q-2P) e^{-2(2-P-Q)y}, \\
\label{eq:delta11}
& & \delta_1 = - 2\alpha_1 - 2 \dfrac{d\alpha_1}{dy} - \tilde{\mathcal{C}} e^{-2(2-P-Q)y},
\end{eqnarray}
where $\mathcal{C}$ denotes the combination of previously defined constants,
\begin{eqnarray}
\tilde{\mathcal{C}} = \dfrac{8\pi\kappa}{3H_0^2} K q^2 \left(\frac{c}{a_0}\right)^{2(P+Q)}.
\end{eqnarray}
It is easy to find solutions of equations (\ref{eq:alpha11})-(\ref{eq:delta11}). We have
\begin{eqnarray}
\label{eq:alphay}
& & \alpha_1 = c_1 + c_2 e^y + \dfrac{\tilde{\mathcal{C}}e^{-2(2-P-Q)y}}{10-4P-4Q}, \\
\label{eq:sigmay}
& & \sigma_1 = c_3 + c_4 e^y + \dfrac{\tilde{\mathcal{C}}(Q-2P)e^{-2(2-P-Q)y}}{(4-2P-2Q)(5-2P-2Q)}, \\
\label{eq:deltay}
& & \delta_1 = - 2c_1 - 4c_2 e^y - \dfrac{2\tilde{\mathcal{C}}e^{-2(2-P-Q)y}}{5-2P-2Q},
\end{eqnarray}
and by using definition (\ref{eq:defy}) we can also find these solution as functions of the conformal time,
\begin{eqnarray}
\label{eq:alpha1tau}
& & \alpha_1 = K_1 + K_2 \tau^{-1} + \mathcal{C}_1\tau^{2(2-P-Q)}, \\
\label{eq:sigma1tau}
& & \sigma_1 = K_3 + K_4 \tau^{-1} + \mathcal{C}_2\tau^{2(2-P-Q)}, \\
\label{eq:delta1tau}
& & \delta_1 = -2K_1 - 4K_2 \tau^{-1} - 4 \mathcal{C}_1\tau^{2(2-P-Q)},
\end{eqnarray}
with constants $\mathcal{C}_1$ and $\mathcal{C}_2$ defined as
\begin{eqnarray}
& & \mathcal{C}_1 = \dfrac{\mathcal{C}}{10-4P-4Q}, \quad \mathcal{C}_2 = \dfrac{\mathcal{C}(Q-2P)}{(4-2P-2Q)(5-2P-2Q)}, \\
& & \mathcal{C} = \dfrac{8\pi\kappa}{3}H_0^{-2(3-P-Q)} K q^{-2(1-P-Q)} c^{2(P+Q)} a_0^{-4}. \nonumber
\end{eqnarray}
Parts given by constants $c_1$-$c_4$ and $K_1$-$K_4$ are homogeneous solutions that correspond to the case without presence of the anisotropic solid remnant. In this case, the parameter $\gamma=\dot{\sigma}/\dot{\alpha}$ is $-\lambda K_3 \tau^{-1}$ up to the first order in parameter $\lambda$, which means that the anisotropy decreases with time, as expected when there is no source of the anisotropy.

\vskip 2mm
In the presence of the anisotropic solid remnant, calculation of the parameter of anisotropy $\gamma=\dot{\sigma}/\dot{\alpha}$ up to the first order in the parameter $\lambda$ yields
\begin{eqnarray}\label{eq:gamma}
\gamma &=&\lambda \left( - K_4 \tau^{-1} + 2(2-P-Q) \mathcal{C}_2 \tau^{2(2-P-Q)} \right) = \\
&=&\lambda \left(-c_4 e^y + \dfrac{\tilde{\mathcal{C}}(Q-2P)}{5-2P-2Q} e^{-2(2-P-Q)y} \right), \nonumber
\end{eqnarray}
which means that if $P+Q<2$, the anisotropy increases. This condition can also be intuitively comprehended by comparing contributions of two types of matter under consideration to the total matter Lagrangian. Since $\Theta_0=\Omega_0\propto a^{-2}$ for the background configuration, we have
\begin{eqnarray}
\mathcal{L}_{\textrm{rad.}} \propto a^{-4}, \quad \lambda f = \lambda K \Theta_0^P \Omega_0^Q \propto a^{-2(P+Q)},
\end{eqnarray}
and it is easy to see that when $P+Q>2$, the contribution of the solid remnant decreases with respect the radiation component, and this is associated with decreasing anisotropy. If we disregard homogeneous parts of solutions (\ref{eq:alpha11})-(\ref{eq:delta11}), i.e., we set $c_4=0$, we can evaluate the change of the anisotropy parameter $\gamma$ from the GUT scale ($y=65.92$) to the $100\textrm{TeV}$ scale ($y=40.59$) as
\begin{eqnarray}
\dfrac{\gamma_{100\textrm{TeV}}}{\gamma_{\textrm{GUT}}} = e^{50.66(2-P-Q)},
\end{eqnarray}
which is very sensitive to the difference of $P+Q$ from factor $2$. In the case with a more general form of the function $f$ describing the solid remnant matter Lagrangian, we can expand such function $f$ into a sum of functions of the form (\ref{eq:PQ}). In such case, the potential growth of anisotropy is dominated by parts of the sum with the highest value of the combination of power coefficients $2-P-Q$. Therefore, the special case studied here serves as a good indicator of the anisotropy for the case with general form of the function $f$.

\vskip 2mm
Observations indicate a small level of anisotropy. This includes CMB data\cite{groen,pullen} as well as more recent quasar and type Ia supernovae data\cite{krishnan1,secrest1,secrest2,luongo,krishnan2,zhai,sorrenti,conville,sah,boubel}. In order to avoid growth of the anisotropy and, at the same time, to study the case with the strongest effect of the anisotropic solid remnant, we focus on the case with $P+Q=2$. We will study the evolution of superhorizon perturbations under this assumption in the following section.

\section{Evolution of superhorizon perturbations}\label{sec:4}

Whereas the standard FLRW background with flat space geometry (\ref{eq:flrw}) accommodates three distinctive types of perturbations, scalar, vector, and tensor, due to the full Euclidean background symmetry, the situation with Bianchi type-I background with residual symmetry in two-dimensional plane (\ref{eq:bianchi}) is different. There are no tensor modes in spacetime considered in our model, and we have only scalars and two-dimensional vectors. Before subtracting nonphysical degrees of freedom due to the diffeomorphism invariance, we have seven scalar and three vector perturbation, where each vector perturbation carries only one degree of freedom\cite{gumru}, and the linearized spacetime perturbations can be parameterized as
\begin{eqnarray}\label{eq:met1}
& & g_{00} = -e^{2\alpha}(1+2\phi), \quad g_{01} = e^{\alpha}\partial_1\chi, \quad g_{0A} = e^{\alpha} (\partial_A B+ e^{\sigma} B_A), \\
& & g_{11} = e^{2\alpha-4\sigma}(1-2\psi), \quad g_{1A} = e^{2\alpha+2\sigma}\partial_1(\partial_A\tilde{B}+\tilde{B}_A), \nonumber\\
& & g_{AB} = e^{2\alpha+2\sigma}(1 - 2\Sigma\delta_{AB} + 2\partial_A\partial_B E + \partial_A E_B + \partial_B E_A). \nonumber
\end{eqnarray}
The three vector perturbations are transverse, $B_{A,A}=0$, $\tilde{B}_{A,A}=0$, $E_{A,A}=0$, and quantities $\phi$, $\chi$, $B$, $\psi$, $\tilde{B}$, $\Sigma$, $E$ parameterize seven scalar degrees of freedom.

\vskip 2mm
It is convenient to decompose perturbations into Fourier modes associated with wavectors $\mathbf{k}=(k_1,k_2,k_3)$. With the FLRW background, we can, for example, set $\mathbf{k}=(k,0,0)$ without any loss of generality, because of the background isotropy, which causes equations for perturbations to be isotropic as well. However, with only the residual isotropy in $\{x^2,x^3\}$-plane, we can set only one of $k_2$ and $k_3$ to zero. Therefore, from now on, we choose $\mathbf{k}=(k_1,k_2,0)$, which simplifies the analysis while keeping the results general. Transversal conditions for vector perturbations then imply that only the third component of vector modes are nonzero, $B_{A,A}=0$ $\Rightarrow$ $k_2B_2=0$ $\Rightarrow$ $B_A = \delta^3_A B_3$.

\vskip 2mm
This choice is useful for comparing the perturbations defined above with the standard scalar, vector, and tensor perturbations defined on the FLRW background\cite{bardeen}. With this convention, we can write components of the perturbed FLRW spacetime metric through Fourier modes of gauge invariant perturbations, which we denote here by bar, as
\begin{eqnarray}\label{eq:met2}
& & g_{00}=-a^2(1+2\overline{\phi}), \quad g_{01}=a^2\frac{k_2}{k}\overline{S}_{+}, \quad g_{02} = -a^2\frac{k_1}{k}\overline{S}_{+}, \quad g_{03} = a^2\overline{S}_{\times}, \\
& & g_{11} = a^2\left(1-2\overline{\psi}+\frac{k_2^2}{k^2}\overline{h}_{+}\right), \quad g_{12} -a^2\frac{k_1k_2}{k^2} \overline{h}_{+}, \quad g_{13} = a^2\frac{k_2}{k}\overline{h}_{\times}, \nonumber\\
& & g_{22} = a^2\left( 1-2\overline{\psi}+\frac{k_1^2}{k^2}\overline{h}_{+} \right), \quad g_{23}=-a^2\frac{k_1}{k}\overline{h}_{\times}, \quad g_{33}=a^2\left(1-2\overline{\psi}-\overline{h}_{+}\right), \nonumber
\end{eqnarray}
where $\overline{\phi}$, $\overline{\psi}$ are two scalar perturbations, $\overline{S}_{+}$, $\overline{S}_{\times}$ are two independent polarizations of vector perturbations, $\overline{h}_{+}$, $\overline{h}_{\times}$ are polarizations of tensor perturbations, and all these perturbations correspond to Fourier modes with wavector $\mathbf{k}=(k_1,k_2,0)$, where we have also simplified the usual notation by skipping wavevectors index indicating Fourier modes, $\overline{\phi}_{\mathbf{k}} \equiv \overline{\phi}$.
In order to distinguish perturbations used here from two-dimensional scalar and two-dimensional vector perturbations defined in (\ref{eq:met1}), we call $\overline{\phi}$ and $\overline{\psi}$ three-dimensional scalar, $\overline{S}_{+}$, $\overline{S}_{\times}$ three-dimensional vector, and $\overline{h}_{+}$, $\overline{h}_{\times}$ three-dimensional tensor perturbations, since they are defined in the standard way with respect to the fully isotropic background with three-dimensional rotational symmetry.

\vskip 2mm
Comparing (\ref{eq:met1}) with (\ref{eq:met2}) is meaningful only in the limit of vanishing background anisotropy, $\sigma\to 0$. If we make this comparison in the limit of the background isotropy, and take into account gauge transformation that sets perturbations $\tilde{B}$, $\Sigma$, $E$ and $E_A$ to zero, we obtain
\begin{eqnarray}\label{eq:transf}
& & \overline{\phi} = \phi - \left[1+\left(\dfrac{1}{\alpha^{\prime}}\right)\right] \Gamma - \dfrac{1}{\alpha^{\prime}} \Gamma^{\prime}, \quad \Gamma = \alpha^{\prime} \dfrac{1}{k^2} \left[-e^{-\alpha}\left(k_1^2\chi+k_2^2B\right) + \dfrac{2k_1^2-k_2^2}{2k^2}\psi^{\prime}\right], \nonumber\\
& & \overline{\psi} = -e^{-\alpha}\alpha^{\prime} \left(\dfrac{k_1^2}{k^2}\chi+\dfrac{k_2^2}{k^2}B\right) + \dfrac{1}{2}\dfrac{k_2^2}{k^2}\psi + \alpha^{\prime} \dfrac{2k_1^2-k_2^2}{2k^4}\psi^{\prime}, \\
& &  \overline{S}_{+} = -i \dfrac{k_1k_2}{k} \left[ e^{-\alpha}\left(\chi-B\right)-\dfrac{2}{k^2}\psi^{\prime}\right], \quad
\overline{S}_{\times} = e^{-\alpha} B_3 - \dfrac{k_1^2}{k^2} \tilde{B}_3^{\prime}, \nonumber\\
& & \overline{h}_{+} = - \dfrac{k_2^2}{k^2} \psi, \quad
\overline{h}_{\times} = -i \dfrac{k_1k_2}{k} \tilde{B}_3. \nonumber
\end{eqnarray}
These transformation relations can be found in\cite{gumru}, where the authors omitted only vector perturbations.
Despite the fact that there are no two-dimensional tensor perturbations, three-dimensional tensor perturbations appear in the limit of vanishing anisotropy through appropriate combinations of perturbations defined on the anisotropic background.
The form of these relations reveal that three-dimensional scalar, vector and tensor perturbations, are given by two-dimensional scalars, and at the same time, $\tilde{B}_3$ enters both three-dimensional vector and tensor perturbations. This means that the three types of perturbations defined on the isotropic background are no longer independent if the background anisotropy is present. Three-dimensional scalar, vector and tensor perturbations can be defined also directly on the anisotropic background\cite{anizpicon,pereira}, where polarization tensors are time dependent and decomposition of perturbations into these three kinds does not commute with the time evolution.

\vskip 2mm
Quantities parameterizing matter perturbations have to be introduced as well. Perturbation of the radiation energy density $\Delta$ and perturbations of its $4$-velocity $\delta u^1$, $\delta u_{\parallel}$ and $\delta u_{\perp A}$ can be defined as
\begin{eqnarray}\label{eq:mat11}
\rho = \overline{\rho}(1+\Delta), \quad u^{\mu} = \delta_0^{\mu} e^{-\alpha} (1-\phi) + \delta_1^{\mu} \delta u^1 + \delta_A^{\mu}\left(\delta u_{\parallel,A}+\delta u_{\perp A}\right),
\end{eqnarray}
where $\overline{\rho}$ is the unperturbed radiation energy density, and the $u^0$ component is given by normalization condition $u_{\mu}u^{\mu}=-1$. For the anisotropic solid remnant fields we have quantities $\delta x$, $A_{\parallel}$, and $A_{\perp A}$ defined as
\begin{eqnarray}\label{eq:mat12}
\Phi^1 = c ( x^1 + \delta x ) , \quad \Phi^A = c ( x^A + A_{\parallel,A} + A_{\perp A} ) .
\end{eqnarray}
Vector perturbations are transverse, $\delta u_{\perp A,A}=0$, $A_{\perp A,A}=0$. Since all the gauge freedom has been used to fix the metric perturbations, we have to take into account all quantities for the matter perturbations defined above.

\vskip 2mm
Observationally relevant modes were superhorizon after inflation, $k\tau \ll 1$, and therefore, we can take the long-wavelength limit, and we will keep only the leading order terms. The full Einstein tensor and stress-energy tensor components can be found in the appendix \ref{app:tensors}. In the long-wavelength limit, the equations reduce to a very manageable form because they partially decouple into several groups. In the first group, we have perturbations $\phi$, $\psi$ and $\Delta$ with the following system of equations
\begin{eqnarray}\label{eq:group1}
\begin{array}{ll}
\Delta + 2\phi + \dfrac{2}{3} \tau \psi^{\prime} = \lambda \left[ \dfrac{2}{3}(K_2-K_4) \psi^{\prime} - \dfrac{2}{3}PI\psi\right], & (00) \\
\Delta + 2\phi - 2 \tau \phi^{\prime} = \lambda\left[-2(K_2-K_4)\phi^{\prime}+2(3P-5)I\psi\right], & (11) \\
\Delta + 2\phi - 2 \tau \left(\phi+\psi\right)^{\prime} - \tau^2\psi^{\prime\prime} =&(\textrm{Tr}{AB}) \\ = \lambda\left[(K_2+2K_4)\phi^{\prime}+(3K_2-2K_4)\psi^{\prime}+2P(P-1)I\psi\right], & \\
\Delta^{\prime} - \dfrac{4}{3}\psi^{\prime} = 0, & (\textrm{rad.})
\end{array}
\end{eqnarray}
where $I = 8\pi\kappa K c^4 a_{*}^{-2} \tau_{*}^2$, we have used also the background solution (\ref{eq:alpha1tau})-(\ref{eq:delta1tau}) with $P+Q=2$, and brackets indicate components of the Einstein field equations with lower indices or conservation laws from which these equations have been derived. First two equations are derived from the $00$ and $11$ components of the Einstein field equations, third one follows from the two-dimensional trace, the sum of the $22$ and $33$ components, and the last equation is given by the conservation law for the radiation component of the matter, $T_{\mu\nu}^{(\textrm{rad.});\mu}u^{\nu}=0$. Similarly, for perturbations $B$, $\delta u_{\parallel}$ and $A_{\parallel}$ we have
\begin{eqnarray}\label{eq:group2}
\begin{array}{ll}
B + e^{2\alpha_0} \delta u_{\parallel} = \lambda \left[ \dfrac{2}{\tau} K_4 B - \dfrac{2}{\tau}(K_2+K_4) e^{2\alpha_0}\delta u_{\parallel} - \dfrac{1}{4}(P-2)I e^{\alpha_0} A_{\parallel}^{\prime} \right], & (0A) \\
B + \tau B^{\prime} = \lambda\left[ \dfrac{1}{\tau}(-2K_2+K_4)B + \dfrac{2}{\tau}(P-2)Ie^{\alpha_0}A_{\parallel}\right], & (\cancel{\textrm{Tr}}AB) \\
B - \tau B^{\prime} + e^{2\alpha_0} \left( \delta u_{\parallel}-\tau\delta u_{\parallel}^{\prime}\right) =&((\textrm{rad.})A) \\ = \lambda\left[ -\dfrac{1}{\tau}K_4 B + \dfrac{1}{\tau}(4K_2+K_4)e^{2\alpha_0}\delta u_{\parallel}-2(K_2+K_4)e^{2\alpha_0}\delta u_{\parallel}^{\prime}\right], &
\end{array}
\end{eqnarray}
with the first two equations derived from the scalar part of $0A$ components and traceless part of $AB$ components of the Einstein field equations, and the third equation is derived from scalar part of the conservation law $T^{(\textrm{rad.});\mu}_{A\mu}=0$. Once the evolution of perturbation $B$ is already known, we have three equations for three perturbations $\chi$, $\delta u^1$, and $\delta x$. They are
\begin{eqnarray}\label{eq:group3}
\begin{array}{ll}
\partial_1\chi + e^{2\alpha_0} \delta u^1 = \lambda \left[ \dfrac{2}{\tau} K_4 \partial_1\chi + \dfrac{2}{\tau}(2K_2+K_4)e^{2\alpha_0} \delta u^1 + \dfrac{1}{2}PI e^{\alpha_0} \delta x^{\prime} \right], & (01) \\
\partial_1 \left[ \chi + B + \tau\left(\chi+B\right)^{\prime}\right] =&\\
= \lambda\left[ \dfrac{1}{\tau}(-2K_2+K_4)\partial_1\chi + \dfrac{1}{\tau} (4K_2+K_4)\partial_1 B - \dfrac{4}{\tau}PI e^{\alpha_0} \delta x \right], & (1A) \\
\partial_1\chi^{\prime} - \dfrac{1}{\tau}\partial_1\chi + e^{2\alpha_0} \left( \delta u^{1\prime} + \dfrac{1}{\tau}\delta u^1\right) = \lambda \dfrac{6}{\tau^2}(2K_2-K_4)\partial_1\chi, & ((\textrm{rad.})1) \\
\end{array}
\end{eqnarray}
where, again, only scalar parts of corresponding equations have been taken, and the last equation is derived from the conservation law, $T^{(\textrm{rad.});\mu}_{1\mu}=0$. Finally, the vector perturbations can be split into two groups. In the first group we have quantities $B_A$, $\delta u_{\perp A}$ and $A_{\perp A}$, and the second group contains only one vector quantity $\tilde{B}_A$. Their equations are
\begin{eqnarray}\label{eq:group4}
& & \begin{array}{ll}
B_A + e^{2\alpha_0} \delta u_{\perp A} =&\\
= \lambda \left[ \dfrac{2}{\tau}K_4 B + \dfrac{1}{\tau}(-K_2+2K_4)e^{2\alpha_0} \delta u_{\perp A} - \dfrac{1}{4}(P-2)I e^{\alpha_0} A_{\perp A}^{\prime} \right], & (0A) \\
B_A + \tau B_A^{\prime} = \lambda \left[ \dfrac{1}{\tau}(-K_2+K_4) B_A + \dfrac{2}{\tau}(P-2)Ie^{\alpha_0} A_{\perp A} \right], & (\cancel{\textrm{Tr}}AB) \\
B_A - \tau B_A^{\prime} + e^{2\alpha_0} \left( \delta u_{\perp A} - \tau \delta u_{\perp A}^{\prime} \right) + \lambda \bigg[ \dfrac{1}{\tau} (K_2-K_4) B_A + & ((\textrm{rad.})A) \\
+ \dfrac{1}{\tau} (3K_2+K_4) e^{2\alpha_0} \delta u_{\perp A} - (K_2+2K_4) e^{2\alpha_0} \delta u_{\perp A}^{\prime}\bigg]=0, &
\end{array} \\
\label{eq:group5}
& & \begin{array}{ll}
\tilde{B}_A^{\prime} + \dfrac{1}{2} \tau \tilde{B}_A^{\prime\prime} = \lambda \left[ \dfrac{1}{\tau}(3K_2+K_4)\tilde{B}_A^{\prime} - \dfrac{1}{\tau}I \tilde{B}_A \right], & (1A)
\end{array}
\end{eqnarray}
with only vector parts taken into account, and the third equation follows from the conservation law $T^{(\textrm{rad.});\mu}_{A\mu}=0$.

\vskip 2mm
When dealing with cosmological perturbations, we will make use of the perturbative approach with respect to parameter $\lambda$ in the same way as in the previous section.
Evolution of other zeroth-order parts is governed by equations (\ref{eq:group1})-(\ref{eq:group5}) without the right-hand sides. Their solutions then can be used as source terms on the right-hand-sides of the derived equations used for the first-order quantities. We also keep only the leading-order terms in the right-hand sides in the limit of the early time. In this way, we find time dependence of superhorizon perturbations up to the first order in parameter $\lambda$ as
\begin{eqnarray}
\label{eq:solution1}
\phi &=&\phi_{(1)} + \phi_{(2)}\zeta^{-1} + \lambda k (K_2-K_4) \phi_{(2)}\zeta^{-2}, \\
\psi &=&\psi_{(1)} - 3\phi_{(2)}\zeta^{-1} - 3\lambda k (K_2-K_4) \phi_{(2)}\zeta^{-2}, \nonumber\\
\Delta &=&-2\phi_{(1)} - 4\phi_{(2)}\zeta^{-1} - 4\lambda k (K_2-K_4) \phi_{(2)}\zeta^{-2}, \nonumber\\
\label{eq:solution2}
\partial_2 B &=&\lambda B_{(1)} \zeta^{\mathcal{Y}}, \\
\partial_2 \delta u_{\parallel} &=&\lambda \dfrac{\mathcal{Y}(1-\mathcal{Y}^2)}{16} \left(\dfrac{k\tau_0}{a_0}\right)^2 B_{(1)} \zeta^{\mathcal{Y}-2}, \nonumber\\
\partial_2 A_{\parallel} &=&\lambda \dfrac{\mathcal{Y}+1}{2} \dfrac{1}{(P-2)I} \dfrac{\tau_0}{a_0} B_{(1)} \zeta^{\mathcal{Y}}, \nonumber\\
\label{eq:solution3}
\partial_1 \chi &=&\chi_{(1)} \zeta^{-1} + \lambda k (2K_2-9K_4) \chi_{(1)} \zeta^{-2} - \lambda B_{(1)} \zeta^{\mathcal{Y}}, \\
\delta u^1 &=&\left(\dfrac{k\tau_0}{a_0}\right)^2 \left[ -\chi_{(1)} \zeta^{-3} + \lambda k (-6K_2+11K_4) \chi_{(1)} \zeta^{-4} + \lambda B_{(1)} \zeta^{\mathcal{Y}-2} \right], \nonumber\\
\delta x &=&- 2 \lambda \dfrac{1}{PI} \dfrac{k\tau_0}{a_0} K_4 \chi_{(1)} \zeta^{-2}, \nonumber\\
\label{eq:solution4}
B_A &=&\lambda B_{(1)A} \zeta^{\mathcal{Y}}, \\
\delta u_{\perp A} &=&\lambda \dfrac{\mathcal{Y}(1-\mathcal{Y}^2)}{16} \left(\dfrac{k\tau_0}{a_0}\right)^2 B_{(1)A} \zeta^{\mathcal{Y}-2}, \nonumber\\
A_{\perp A} &=&\lambda \dfrac{\mathcal{Y}+1}{2} \dfrac{1}{(P-2)I} \dfrac{\tau_0}{a_0} B_{(1)A} \zeta^{\mathcal{Y}}, \nonumber\\
\label{eq:solution5}
\partial_1 \tilde{B}_A &=&\tilde{B}_{(1)A} + \tilde{B}_{(2)A} \zeta^{-1} - \lambda k (3K_2+K_4) \tilde{B}_{(2)A} \zeta^{-2},
\end{eqnarray}
where constant $\mathcal{Y}$ satisfies the algebraic equation $\mathcal{Y}(\mathcal{Y}+1)(\mathcal{Y}-3)=16$, $\zeta = k\tau$ is a dimensionless variable measuring time in multiples of conformal time at which the given mode crosses the horizon, because at the time of horizon crossing $\zeta=1$, and some quantities are expressed in terms of their derivatives in such way they appear in the definitions (\ref{eq:met1}), (\ref{eq:mat11}) and (\ref{eq:mat12}). Quantities with lower indices in brackets are integration constants.
For scalar perturbations we have $\phi_{(1)}$, $\psi_{(1)}$, $\psi_{(2)}$, $B_{(1)}$ and $\chi_{(1)}$, and $B_{(1)A}$, $\tilde{B}_{(1)A}$ and $\tilde{B}_{(2)A}$ for the two-dimensional vector perturbations.

\vskip 2mm
We have used four equations (\ref{eq:group1}) for three quantities $\phi$, $\psi$ and $\Delta$, but the solution (\ref{eq:solution1}) satisfies all of them, which means that, at least under all the considered approximations, only three of the four equations are independent. There is also a very noticeable similarity between the two sets of solutions (\ref{eq:solution2}) and (\ref{eq:solution4}) despite the differences between the right-hand sides of (\ref{eq:group2}) and (\ref{eq:group4}). The reason is that these quantities vanish in the zeroth order in the parameter $\lambda$.
The cubic equation for $\mathcal{Y}$ originates from the condition of the system of three equation for $B$, $\delta u_{\parallel}$ and $A_{\parallel}$ (\ref{eq:group2}), as well as for $B_A$, $\delta u_{\perp A}$ and $A_{\perp A}$ (\ref{eq:group4}), to have non-zero solutions for these perturbations. At the same time, to have non-zero solutions, they cannot be independent. Three perturbations (\ref{eq:solution2}) are given by one constant $B_{(1)}$, and the same is true for (\ref{eq:solution4}) given by $B_{(1)A}$.

\vskip 2mm
Since the constant $\mathcal{Y}$ appearing in (\ref{eq:solution2})-(\ref{eq:solution4}) satisfies a cubic equation, it may have three different values, $\mathcal{Y}_1$-$\mathcal{Y}_3$. It turns out that only one solution $\mathcal{Y}_1$ is real, and other two $\mathcal{Y}_2$ and $\mathcal{Y}_3$ are complex,
\begin{eqnarray}
& & \mathcal{Y}_1 = \dfrac{1}{3}\left(2+\mathcal{Y}_0+\dfrac{13}{\mathcal{Y}_0}\right), \quad \mathcal{Y}_0=\left(251-6\sqrt{3\cdot 563}\right)^{1/3}\approx 1.64,\\
& & \mathcal{Y}_{2,3} = \dfrac{1}{6}\left(4-\mathcal{Y}_0-\dfrac{13}{\mathcal{Y}_0}\right) \pm i\dfrac{\sqrt{3}}{6}\left(\mathcal{Y}_0-\dfrac{13}{\mathcal{Y}_0}\right),
\end{eqnarray}
with the numerical values $\mathcal{Y}_1 \approx 3.85$ and $\mathcal{Y}_{2,3} \approx -0.928 \mp i 1.81$. The consequence of the imaginary power would be an existence of modes containing functions of the form $\cos(\textrm{Im}\{\mathcal{Y}_{2,3}\}\cdot\ln\zeta)$. Wavefronts corresponding to such modes would then be given by
\begin{eqnarray}
\textrm{Im}\{\mathcal{Y}_{2,3}\} \cdot \ln (k|\tau|) - k_i x^i = \textrm{const.},
\end{eqnarray}
implying speed of propagation of such wavefronts $|d\vec{x}/d\tau| = \textrm{Im}\{\mathcal{Y}_{2,3}\}\zeta^{-1}$. This means that there would not only be superluminal superhorizon modes, but their sound speed would also approach infinity in the limit of very large wavelengths. It is then reasonable to consider only the pure real solution $\mathcal{Y}_1$ to be physically relevant.

\vskip 2mm
However, this choice also leads to some problems. Since $\mathcal{Y}_1\approx 3.85$, there are growing modes in (\ref{eq:solution2})-(\ref{eq:solution4}), indicating instability. These growing modes are negligible in the superhorizon limit, $\zeta\to 0$, but after long enough time such modes would grow in such way that the perturbation theory breaks down. The simplest solution of this problem is to consider the period during which the model under consideration holds to be short. Assuming all integration constants are of the same order, the growing modes do not exceed the constant modes if the scale factor at the end of the reheating period $a_{\textrm{reh.}}$ is related to its value at the end of the inflation $a_{\textrm{inf.}}$ by the relation
\begin{eqnarray}\label{eq:stabilimit}
\dfrac{a_{\textrm{reh.}}}{a_{\textrm{inf.}}} < \lambda^{-1/\mathcal{Y}_1}.
\end{eqnarray}
For example for $\lambda = 10^{-3}$ we have the factor $\lambda^{-1/\mathcal{Y}_1}\approx 6$, which in terms of the cosmic time yields $t_{\textrm{reh.}} < 36 t_{\textrm{inf.}}$, since $a\propto \sqrt{t}$. Therefore, there can be a long enough period during which the instability does not cause significant problems, after which the anisotropic solid remnant, the source of the instability, dissipates.

\vskip 2mm
An exciting possibility is that the dissipation of the solid remnant would be caused by effects of its instability due to non-linear effects. This had been studied in inflationary model with two fields\cite{prokopec}, as well as in single field models\cite{amin1}, through both linearized theory\cite{turzynski,fonseca} and numerical methods\cite{amin2,amin3,morgante}. This would require a more detailed description of the solid remnant fields, as opposed to the simplified matter Lagrangian (\ref{eq:segments}), and also non-perturbative approach. For now, we leave this for future work.

\vskip 2mm
In order to obtain results in terms of the standard scalar, vector, and tensor perturbations (\ref{eq:met2}), which are very well understood in the context of the observational consequences, we have to insert results (\ref{eq:solution1})-(\ref{eq:solution5}) into the transformation relations (\ref{eq:transf}). In this way, we obtain
\begin{eqnarray}
\label{eq:phipsi}
\overline{\phi} &=&\phi_{(1)} + \phi_{(2)} \zeta^{-1} + \Gamma_{(1)} \zeta^{-3} + \\
& & + \lambda k (K_2-K_4) \phi_{(2)} \zeta^{-2} + 2\lambda \Gamma_{(2)} \zeta^{-4} - \lambda \mathcal{Y}_1 \Gamma_{(3)} \zeta^{\mathcal{Y}_1-2}, \nonumber\\
\overline{\psi} &=&\dfrac{1}{2}\dfrac{k_2^2}{k^2}\psi_{(1)} - \dfrac{3}{2}\dfrac{k_2^2}{k^2} \phi_{(2)} \zeta^{-1} + \Gamma_{(1)} \zeta^{-3} - \nonumber\\
& & -\dfrac{3}{2} \lambda \dfrac{k_2^2}{k} (K_2-K_4) \phi_{(2)} \zeta^{-2} + \lambda \Gamma_{(2)} \zeta^{-4} + \lambda \Gamma_{(3)} \zeta^{\mathcal{Y}_1-2}, \nonumber\\
& & \Gamma_{(1)} = -\dfrac{k\tau_0}{a_0} i \dfrac{k_1}{k} \chi_{(1)} + 3 \dfrac{2k_1^2-k_2^2}{2k^2} \phi_{(2)}, \nonumber\\
& & \Gamma_{(2)} = -\dfrac{k\tau_0}{a_0} i k_1 (2K_2-9K_4) \chi_{(1)} + 3 \dfrac{2k_1^2-k_2^2}{k} (K_2-K_4) \phi_{(2)}, \nonumber\\
& & \Gamma_{(3)} = \dfrac{k\tau_0}{a_0} i \dfrac{k_2}{k} \left(\dfrac{k_1^2}{k_2^2}-1\right) B_{(1)}, \nonumber\\
\label{eq:ss}
\overline{S}_{+} &=&\dfrac{k_1 k_2}{k^2} \Gamma_{(4)} \zeta^{-2} + \lambda \Gamma_{(5)} \zeta^{-3} - 2 \lambda \dfrac{k\tau_0}{a_0} \dfrac{k_1}{k} B_{(1)} \zeta^{\mathcal{Y}_1-1}, \\
\overline{S}_{\times} &=&\dfrac{k_1^2}{k} \tilde{B}_{(2)3} \zeta^{-2} - 2\lambda k_1^2(3K_2+K_4)B_{(2)3} \zeta^{-3} + \lambda \dfrac{k\tau_0}{a_0} B_{(1)3} \zeta^{\mathcal{Y}_1-1}, \nonumber\\
& & \Gamma_{(4)} = \dfrac{k\tau_0}{a_0} \dfrac{k}{k_1} \chi_{(1)} + 6i\phi_{(2)}, \nonumber\\
& & \Gamma_{(5)} = \dfrac{k\tau_0}{a_0} k_2 (2K_2-9K_4) \chi_{(1)} + 12i \dfrac{k_1 k_2}{k} (K_2-K_4) \phi_{(2)}, \nonumber\\
\label{eq:hh}
\overline{h}_{+} &=&- \dfrac{k_2^2}{k^2} \psi_{(1)} + \dfrac{3}{2} \dfrac{k_2^2}{k^2} \phi_{(2)} \zeta^{-1} + 3 \lambda \dfrac{k_2^2}{k} (K_2-K_4) \phi_{(2)} \zeta^{-2}, \\
\overline{h}_{\times} &=&\dfrac{k_2}{k} \tilde{B}_{(1)3} + \dfrac{k_2}{k} \tilde{B}_{(2)3}\zeta^{-1} - \lambda k_2 (3K_2+K_4) \tilde{B}_{(2)3} \zeta^{-2}. \nonumber
\end{eqnarray}
This allows us to discuss differences from the simplest early Universe model - isotropic and flat Universe filled with radiation perfect fluid. In this model, the scalar perturbations $\overline{\phi}$ and $\overline{\psi}$ are equal, and their modes evolve as
\begin{eqnarray}
\overline{\phi} = \overline{\psi} = c_{(1)} \dfrac{1}{u^2}\left(\dfrac{\sin u}{u}-\cos u\right) + c_{(2)} \dfrac{1}{u^2}\left(\dfrac{\cos u}{u}+\sin u\right),
\end{eqnarray}
where $u=\zeta/\sqrt{3}$, both vector polarizations decrease as $a^{-2}$, and the evolution of two tensor polarizations is given by
\begin{eqnarray}
\overline{h}_{+} = c_{(4,+)} \dfrac{\sin\zeta}{\zeta} + c_{(5,+)} \dfrac{\cos\zeta}{\zeta}, \quad \overline{h}_{\times} = c_{(4,\times)} \dfrac{\sin\zeta}{\zeta} + c_{(5,\times)} \dfrac{\cos\zeta}{\zeta}.
\end{eqnarray}
The superhorizon modes then can be approximated as
\begin{eqnarray}\label{eq:standard}
& & \overline{\phi} = \overline{\psi} = \dfrac{1}{3}c_{(1)} + \dfrac{\sqrt{3}}{2}c_{(2)}\zeta^{-1} + 3\sqrt{3}c_{(2)}\zeta^{-3}, \\
& & \overline{S}_{+} = c_{(3,+)} \zeta^{-2}, \quad \overline{S}_{\times} = c_{(3,\times)} \zeta^{-2}, \nonumber\\
& & \overline{h}_{+} = c_{(4,+)} + c_{(5,+)} \zeta^{-1}, \quad \overline{h}_{\times} = c_{(4,\times)} + c_{(5,\times)} \zeta^{-1}. \nonumber
\end{eqnarray}
We can see that if we set the parameter $\lambda$ to zero, which means no presence of the anisotropic solid remnant, the evolution of vector and tensor modes (\ref{eq:ss}) and (\ref{eq:hh}) coincide with the relation above that are valid in the isotropic flat Universe with radiation. This is not true with scalar perturbations (\ref{eq:phipsi}) where $\overline{\phi}$ and $\overline{\psi}$ differ from each other even without the presence of the solid remnant, and the difference between them depends on the direction in which their modes propagate, as indicated by the wavenumber components appearing in the form of $\overline{\psi}$. From the technical point of view, this discrepancy originates from the form of $\Gamma$ defined in (\ref{eq:transf}), and it should be present also in other models with anisotropic expansion, but in the leading order of the superhorizon approximation, they coincide, $\overline{\phi}\approx\overline{\psi}\approx\Gamma_{(1)}\zeta^{-3}$.

\vskip 2mm
Effects of the anisotropy are given by parts of modes (\ref{eq:phipsi})-(\ref{eq:hh}) proportional to the parameter $\lambda$. There are two types of such effects. Terms with constants $K_2$ and $K_4$ correspond to the part of the anisotropic expansion without the presence of the solid remnant, because they are associated with only the homogeneous solution of the background equations (\ref{eq:alpha1})-(\ref{eq:delta1}). This is the effect of purely metric anisotropy. Other terms linear in $\lambda$ are then the result of the presence of the solid remnant. Behavior of all parts can be summarized in the table \ref{tab:cases}.
\begin{table}[!htb]
\begin{center}
\begin{tabular}{ | l | c | c | c | }
\hline\hline
modes & parts with $\lambda=0$ & metric anisotropy & solid remnant effect \\
\hline
scalar: $\overline{\phi}$, $\overline{\psi}$ & $0$, $-1$, $-3$ & $-2$, $-4$ & $\mathcal{Y}_1-2\approx 1.85$ \\
vector: $\overline{S}_{+}$, $\overline{S}_{\times}$ & $-2$ & $-3$ & $\mathcal{Y}_1-1\approx 2.85$ \\
tensor: $\overline{h}_{+}$, $\overline{h}_{\times}$ & $0$, $-1$ & $-3$ & \\
\hline\hline
\end{tabular}
\setlength{\abovecaptionskip}{4pt plus 0pt minus 0pt}
\captionsetup{width=12.2cm}
\caption{\label{tab:cases} {\footnotesize Numbers in the table indicate the power of $\zeta$ at each part of the superhorizon solutions (\ref{eq:phipsi})-(\ref{eq:hh}). The first column coincides with the standard behavior in the isotropic flat Universe with radiation (\ref{eq:standard}), the second column describes effect of the metric anisotropy, and the third column reflects presence of the anisotropic solid remnant.}}
\end{center}
\end{table}
As we can see, the solid remnant causes growth of scalar and vector modes, while tensor perturbations have only decaying modes.

\section{Observational consequences}\label{sec:5}

The presence of the anisotropic solid remnant considerably changes the evolution of the superhorizon perturbations in comparison with a more standard model of the Universe filled with only radiation. There are growing superhorizon modes of scalar and vector perturbations, while no growing modes of tensor perturbation appear. Therefore, the primordial tensor-to-scalar ratio $r$ associated with perturbations generated during the cosmic inflation changes, and the same is true also for the vector-to-scalar ratio $s$. This is radically different from the standard scenario, where as long as the modes under consideration remain in the superhorizom regime, the absence of any growing modes preserves the tensor-to-scalar ratio from the end of the inflation to the beginning of the standard $\Lambda$CDM expansion, and vector-to-scalar ratio is being suppressed as $s\propto a^{-4}$, taking away any observational relevance of the vector perturbations.

\vskip 2mm
The tensor-to-scalar ratio decreases, while the behavior of vector-to-scalar ratio is more complicated. If we disregard decaying modes and assume that all remaining constants introduced in (\ref{eq:solution1})-(\ref{eq:solution5}) are of the same order, we can qualitatively describe the behavior of the three types of perturbations as functions of the scale factor in the following way:
\begin{eqnarray}\label{eq:behavior}
\dfrac{\textrm{scalar}}{\textrm{scalar}_{\textrm{inf.}}} & \approx & 1 + \lambda \mathcal{O}(1) \left[\left(\dfrac{a}{a_{\textrm{inf.}}}\right)^{\mathcal{Y}_1-2}-1\right], \\
\dfrac{\textrm{vector}}{\textrm{vector}_{\textrm{inf.}}} & \approx & \left(\dfrac{a}{a_{\textrm{inf.}}}\right)^{-2} + \lambda\mathcal{O}(1)\left[\left(\dfrac{a}{a_{\textrm{inf.}}}\right)^{\mathcal{Y}_1-1}-1\right], \nonumber\\
\dfrac{\textrm{tensor}}{\textrm{tensor}_{\textrm{inf.}}} & \approx & 1, \nonumber
\end{eqnarray}
where $\mathcal{O}(1)$ denotes a factor of order of unity. To simplify the analysis even more and to make the following discussion clearer, we may set all of these factors equal to one, $\mathcal{O}(1)=1$. In this way, we may qualitatively estimate the values of tensor-to-scalar and vector-to-scalar ratios at end of the studied era as
\begin{eqnarray}\label{eq:ratios}
\widetilde{r}=\dfrac{r_{\textrm{reh.}}}{r_{\textrm{inf.}}} \sim \dfrac{1}{\left[1+\lambda\left(\widetilde{a}^{\mathcal{Y}_1-2}-1\right)\right]^2}, \quad
\widetilde{s}=\dfrac{s_{\textrm{reh.}}}{s_{\textrm{inf.}}} \sim \left[\dfrac{\widetilde{a}^{-2}+\lambda\left(\widetilde{a}^{\mathcal{Y}_1-1}-1\right)}{1+\lambda\left(\widetilde{a}^{\mathcal{Y}_1-2}-1\right)}\right]^2,
\end{eqnarray}
\begin{figure}[!htb]
\centering
\sbox0{
\includegraphics[scale=1]{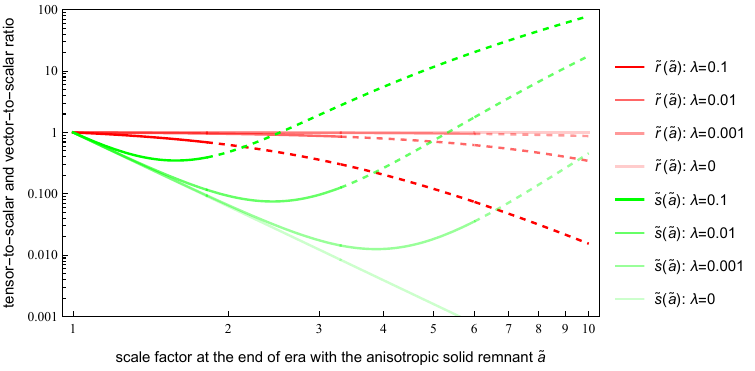}
}
\begin{minipage}{\wd0}
\usebox0
\linespread{1}
\setlength{\abovecaptionskip}{-8pt plus 0pt minus 0pt}
\caption{{\footnotesize Qualitative behavior of the tensor-to-scalar (red) and vector-to-scalar (green) ratio (\ref{eq:ratios}) as functions of the scale factor at the end of era with presence of the anisotropic solid remnant for various values of the parameter $\lambda$. Dashed lines indicate regime when the condition (\ref{eq:stabilimit}) no longer holds.}}
\label{fig:obraz}
\end{minipage}
\end{figure}
where $\widetilde{a}=a_{\textrm{reh.}}/a_{\textrm{inf.}}$. The dependence of these quantities on $\widetilde{a}$ for various values of $\lambda$ is plotted in figure \ref{fig:obraz}. If we want to avoid the breakdown of the linearized perturbation theory due to growth of modes, the relation for the value of the scale factor $a_{\textrm{reh.}}$ at the end of era when the model studied in this paper holds (\ref{eq:stabilimit}) has to be satisfied, $\widetilde{a}<\lambda^{-1/\mathcal{Y}_1}$. Regime in which this condition holds is indicated by solid lines, and dashed lines follow formulas (\ref{eq:ratios}) for $\widetilde{a}>\lambda^{-1/\mathcal{Y}_1}$ when the linearized perturbation theory does not hold, and the used formulas are no longer valid. We plot the dashed lines, because it is reasonable to assume that they at least indicate potential nonlinear trends. As we can see, the tensor-to-scalar ratio is suppressed, and the vector-to-scalar ratio is enhanced in comparison with the standard scenario, when only radiation is present. In the standard case, $\widetilde{r}=1$, and $\widetilde{s}=\widetilde{a}^{-4}$. In general, we obtain stronger effects with the larger values of the parameter $\lambda$, which is in agreement with expectations.

\vskip 2mm
If the value of $\lambda$ is small, but not by many orders of magnitude smaller than unity, the tensor-to-scalar and vector-to-scalar ratios change, but they remain of the same order of magnitude. Observations indicate small values of the tensor-to-scalar ratio, $r<0.032$\cite{tristram} with recent hints towards even much smaller values\cite{wang}, and the model studied in this paper predicts its decrease. The solid inflation model\cite{endlich} with the matter Lagrangian (\ref{eq:solidinflation}) has a wide parameter space which accommodates regions compatible with the observational data. However, if we restrict ourselves to the special case of the solid inflation with the matter Lagrangian of the form as in (\ref{eq:segments}), $\mathcal{L}_{\textrm{m}}\propto X^{\epsilon}$, the theoretical prediction of the tensor-to-scalar ratio becomes larger than desired. In this case, the predicted scalar spectral index $n_{\textrm{S}}$ and tensor-to-scalar ratio are
\begin{eqnarray}
n_{\textrm{S}} = 1 + \dfrac{2}{3}(1+2\epsilon)\epsilon, \quad r_{\textrm{inf.}} = \dfrac{16}{9\sqrt{3}}(1+2\epsilon)^{5/2}\epsilon.
\end{eqnarray}
By using these relations, we find that the value of scalar spectral index $n_{\textrm{S}}=0.965$\cite{planck2} corresponds to the value of tensor-to-scalar ratio $r_{\textrm{inf.}}=0.045$. This value decreases during the era with presence of the solid remnant. By using relation in (\ref{eq:ratios}) and assuming that the studied era lasts until the condition (\ref{eq:stabilimit}) holds, we find that the final value of the tensor-to-scalar ratio reduces to $r_{\textrm{reh.}}=0.038$ for $\lambda=0.01$, and $r_{\textrm{reh.}}=0.031$ for $\lambda=0.1$. Since relations (\ref{eq:ratios}) have been derived with some simplifications, these numbers represent only qualitative estimates, but they demonstrate that the presence of the anisotropic solid remnant improves compatibility of the simplified version of the solid inflation, $\mathcal{L}_{\textrm{m}}\propto X^{\epsilon}$, with the observational restrictions. Of course, this compatibility can be improved also by using the full version of the solid inflation model (\ref{eq:solidinflation}), but the simplified solid inflation model predicts much smaller primordial non-Gaussianity than its full version, which is preferred by observations\cite{planck3,mccarthy,cagliari,jung,fabbian}. The reason is that non-linearity parameters are proportional to $F_Y$ or $F_Z$ in the leading order\cite{endlich}. 

\vskip 2mm
We also obtain another set of interesting results in the limit $\lambda\to 0$, because very small values of $\lambda$ allow the existence of a prolonged period during which the condition of validity of the linearized perturbation theory (\ref{eq:stabilimit}) holds. Using this condition, we can rewrite (\ref{eq:ratios}) to estimate the maximal change of $\widetilde{r}$ and $\widetilde{s}$ before this condition breaks. This yields
\begin{eqnarray}
\widetilde{r} > \widetilde{r}_{\textrm{min}} \sim \dfrac{1}{\left(1+\lambda^{2/\mathcal{Y}_1}-\lambda\right)^2}, \quad \dfrac{\widetilde{s}}{\widetilde{a}^{-4}} < \dfrac{\widetilde{s}_{\textrm{max}}}{\widetilde{a}^{-4}} \sim \left(\dfrac{1+\lambda^{-1/\mathcal{Y}_1}-\lambda^{1-2/\mathcal{Y}_1}}{1+\lambda^{2/\mathcal{Y}_1}-\lambda}\right)^2,
\end{eqnarray}
where in the second relation, the vector-to-scalar ratio is compared with the unmodified behavior, $\widetilde{s}=\widetilde{a}^{-4}$. In the limit $\lambda\to 0$, the leading-order behavior of the limiting values is $\widetilde{r}_{\textrm{min}}\sim 1$ and $\widetilde{s}_{\textrm{max}}/\widetilde{a}^{-4}\sim\lambda^{-2/\mathcal{Y}_1}$. The overall tensor-to-scalar ratio then remains practically unchanged, while the vector-to-scalar ratio may be considerably enhanced. If we place the model studied in this paper between the energy scales of GUT and $100\textrm{TeV}$, as we discussed in section \ref{sec:2}, we obtain $\widetilde{a}\sim 10^{11}$, $\lambda\sim 10^{-11 \mathcal{Y}_1}$, and $\widetilde{s}/\widetilde{a}^{-4}\sim 10^{22}$, which represents a result many orders of magnitude different from the result of the standard scenario, where $\widetilde{s}\sim 10^{-44}$. This means that the vector perturbations are still being suppressed in the course of the expansion of the Universe, $\widetilde{s}\sim\widetilde{a}^{-4}\lambda^{-2/\mathcal{Y}_1}\sim\lambda^{2/\mathcal{Y}_1}\sim \widetilde{a}^{-2}$, but we obtain a novel dependence on the scale factor,
\begin{eqnarray}
\dfrac{\textrm{vector}_{\textrm{reh.}}}{\textrm{vector}_{\textrm{inf.}}} \sim \left(\dfrac{a_{\textrm{reh.}}}{a_{\textrm{inf.}}}\right)^{-1},
\end{eqnarray}
as opposed to the standard-case relation: $\textrm{vector}\propto a^{-2}$. Of course, this modification is valid only in the limit of very small $\lambda$ and the longest possible era with the presence of the anisotropic solid remnant before the breakdown of the perturbation theory.

\section{Summary and outlook}\label{sec:6}

We have analyzed effects of presence of a small amount of so-called anisotropic solid remnant parameterized by a triplet of fields $\Phi^{1}$, $\Phi^{A}$ in the early Universe dominated by radiation. These fields enter the matter Lagrangian (\ref{eq:segments}) through quantities (\ref{eq:thetaomega}) in such way that the full global Euclidean symmetry of the solid inflation\cite{endlich} breaks down to $SO(2)\ltimes T(3)$ symmetry causing anisotropy in one spatial direction. We determined evolution of the anisotropy (\ref{eq:gamma}) under the assumption of the solid remnant matter Lagrangian being of the form (\ref{eq:PQ}), which serves as a good indicator of potential growth of anisotropy with respect to the most general form of the matter Lagrangian (\ref{eq:segments}).

\vskip 2mm
We have focused on the limiting case in which anisotropy remains present but at the smallest possible level. In this way, we avoid too strong anisotropy which would contradict observations, and at the same time, we obtain the strongest possible effects on cosmological perturbations. Since we studied post-inflationary period, we could restrict ourselves to the superhorizon limit. We found, that the presence of the anisotropic solid remnant does not affect tensor perturbations, while scalar and vector perturbations obtain some growing modes. This leads to interesting consequences in terms of tensor-to-scalar and vector-to-scalar ratios. We showed that if the parameter $\lambda$ measuring the contribution of the solid remnant to the total energy density is of order $0.1$ or $0.01$, the tensor-to-scalar ratio is being suppressed, however, it remains of the same order of magnitude. This helps to improve observational consistency of the special case of  solid inflation model (\ref{eq:solidinflation}) with matter Lagrangian of the form $\mathcal{L}_{\textrm{m}}\propto X^{\epsilon}$ which implies suppression of the primordial non-Gaussianity. We found also increase of the vector-to-scalar ratio. This is not surprising, since vector perturbations are much more relevant in the solid inflation model and its generalizations. We also found, that in the limit $\lambda\to 0$ the vector perturbations decrease only as $a^{-1}$ as opposed to the standard proportionality to $a^{-2}$, if the era with presence of the solid remnant is prolonged until the linearized perturbation theory breaks down due to growth of superhorizon modes.

\vskip 2mm
Superhorizon perturbations are not conserved even in the solid inflation model, where their time dependence is small, given by slow-roll parameters. In our previous work, we studied a post-inflationary model, where solid matter fields are the only matter component, predicting a stronger growth of superhorizon perturbations for some values of parameters of the model\cite{meszaros}. Growth of modes is problematic because of instability. However, such instability may lead to very interesting effects studied in the literature\cite{prokopec,amin1,turzynski,fonseca,amin2,amin3,morgante}. Non-linear effects may cause stabilization, and they can produce gravitational waves\cite{khleb,easther} or be an origin of cold dark matter particles\cite{eroncel}. The model studied in this paper, inspired by the solid inflation model, represents a novel approach to the anisotropic expansion of the Universe, and we believe that future works including also non-linear effects can bring even more interesting results with potential to address current problems in cosmology.

\section*{Acknowledgements}
The work was supported by grants VEGA 1/0719/23, VEGA 1/0025/23, VEGA 1/0565/25, UK/1362/2025, and Ministry of Education contract No. 0466/2022.

\appendix
\renewcommand{\thesection}{\Alph{section}}

\section{Full Einstein and stress-energy tensors}\label{app:tensors}

Here we present the components of the full perturbed part of the Einstein tensor corresponding to (\ref{eq:met1}) with the gauge condition $\tilde{B}=E=E_3=\Sigma=0$, the full perturbed part of the stress-energy tensor and the conservation laws used in this work. The Einstein tensor perturbation is
\begin{eqnarray}\label{eq:Einstein_Tensor}
    \delta G_{00}  &=&-e^{-\alpha+4 \sigma}\left(2 \alpha^{\prime}+2 \sigma^{\prime}\right) \partial_1^2 \chi 
 -e^{-\alpha-2 \sigma}\left(2 \alpha^{\prime}-\sigma^{\prime}\right) \partial_A \partial_A B +
 \\
 & &+e^{-2 \sigma} \partial_A \partial_A \psi-\left(2 \alpha^{\prime}+2 \sigma^{\prime}\right) \psi^{\prime},\nonumber\\
 \delta G_{01} &=&\left(2 \alpha^{\prime}+2 \sigma^{\prime}\right) \partial_1 \phi - e^{-\alpha}\left(2 \alpha^{\prime \prime}+2 \sigma^{\prime \prime}+\alpha^{\prime 2}+4 \alpha^{\prime} \sigma^{\prime}+3 \sigma^{\prime 2}\right) \partial_1 \chi -
 \nonumber\\
 & &-\frac{1}{2} e^{-\alpha-2 \sigma} \partial_1 \partial_A \partial_A \chi+
 \frac{1}{2} e^{-\alpha-2 \sigma} \partial_1 \partial_A \partial_A B,\nonumber\\
  \delta G_{0A}&=&\left(2 \alpha^{\prime}-\sigma^{\prime}\right) \partial_A \phi+\frac{1}{2} e^{-\alpha+4 \sigma} \partial_1^2 \partial_{A} \chi - \nonumber\\ & & -e^{-\alpha}\left(2 \alpha^{\prime \prime}-
 \sigma^{\prime \prime}+\alpha^{\prime 2}-2 \alpha^{\prime} \sigma^{\prime}+3 \sigma^{\prime 2}\right) \partial_A B - \nonumber\\
 & &-\frac{1}{2} e^{-\alpha+4 \sigma} \partial_1^2 \partial_A B- 
 3 \sigma^{\prime} \partial_A \psi+\partial_A \psi^{\prime}- \nonumber\\ & & -e^{-\alpha+\sigma}\left(2 \alpha^{\prime \prime}-\sigma^{\prime \prime}+\alpha^{\prime 2}-2 \alpha^{\prime} \sigma^{\prime}+3 \sigma^{\prime 2}\right) B_A - \nonumber\\
 & &-\frac{1}{2} e^{-\alpha+5 \sigma} \partial_1^2 B_A
 -\frac{1}{2} e^{-\alpha-\sigma} \partial_B \partial_B B_A+ 
 \frac{1}{2} e^{6 \sigma} \partial_1^2 \widetilde{B}_A^{\prime}, \nonumber \\
 \delta G_{11} &=&
e^{-4 \sigma}\left(4 \alpha^{\prime \prime}+4 \sigma^{\prime \prime}+2 \alpha^{\prime 2}+8 \alpha^{\prime} \sigma^{\prime}+6 \sigma^{\prime 2}\right) \phi + \nonumber\\
& & +e^{-6 \sigma} \partial_A \partial_A \phi+e^{-4 \sigma}\left(2 \alpha^{\prime}+2 \sigma^{\prime}\right) \phi^{\prime} + \nonumber\\ 
& &+e^{-4 \sigma}\left(4 \alpha^{\prime \prime}+4 \sigma^{\prime \prime}+2 \alpha^{\prime 2}+8 \alpha^{\prime} \sigma^{\prime}+6 \sigma^{\prime 2}\right) \psi + \nonumber\\ & & + e^{-\alpha-6\sigma}\left(\alpha^{\prime}+\sigma^{\prime}\right) \partial_A \partial_A B + e^{-\alpha-6 \sigma} \partial_A \partial_A B^{\prime},\nonumber\\
\delta G_{1 A}&=&-\partial_1 \partial_A \phi 
 -\frac{1}{2} e^{-\alpha}\left(\alpha^{\prime}-2 \sigma^{\prime}\right) \partial_1 \partial_A \chi- \nonumber\\ & & -\frac{1}{2} e^{-\alpha} \partial_1 \partial_A \chi^{\prime} 
 -\frac{1}{2} e^{-\alpha}\left(\alpha^{\prime}+4 \sigma^{\prime}\right) \partial_1 \partial_A B - \nonumber\\
& &-\frac{1}{2} e^{-\alpha} \partial_1 \partial_A B^{\prime}-\frac{1}{2} e^{-\alpha+\sigma}\left(\alpha^{\prime}+5 \sigma^{\prime}\right) \partial_1 B_A-\frac{1}{2} e^{-\alpha+\sigma} \partial_1 B_A^{\prime} - 
\nonumber\\
& &-e^{2 \sigma}\left(2 \alpha^{\prime \prime}-\sigma^{\prime \prime}+\alpha^{\prime 2}-2 \alpha^{\prime} \sigma^{\prime}+3 \sigma^{\prime 2}\right) \partial_1 \widetilde{B}_A - \nonumber\\  
& &-\frac{1}{2} \partial_1 \partial_B \partial_B \widetilde{B}_A+e^{2 \sigma}\left(\alpha^{\prime}+3 \sigma^{\prime}\right) \partial_1 \widetilde{B}_A^{\prime}+\frac{1}{2} e^{2 \sigma} \partial_1 \tilde{B}_A^{\prime \prime},\nonumber\\
\delta G_{A B}&=&\Big[ 
 e^{2 \sigma}\left(4 \alpha^{\prime \prime}-2 \sigma^{\prime \prime}+2 \alpha^{\prime 2}-4 \alpha^{\prime} \sigma^{\prime}+6 \sigma^{\prime 2}\right) \phi
 +e^{6 \sigma} \partial_1^2 \phi+\partial_C \partial_C \phi + \nonumber\\ & & + e^{2 \sigma}\left(2 \alpha^{\prime}-\sigma^{\prime}\right) \phi^{\prime} + e^{-\alpha+6 \sigma}\left(\alpha^{\prime}+\sigma^{\prime}\right) \partial_1^2 \chi+e^{-\alpha+6 \sigma} \partial_1^2 \chi^{\prime} + \nonumber\\ & & + e^{-\alpha}\left(\alpha^{\prime}-2 \sigma^{\prime}\right) \partial_C \partial_C B+e^{-\alpha} \partial_C \partial_C B^{\prime} - \nonumber\\
& & -\partial_C \partial_C \psi+e^{2 \sigma}\left(2 \alpha^{\prime}-3 \sigma^{\prime}\right) \psi^{\prime}+e^{2 \sigma} \psi^{\prime \prime}\Big] \delta_{A B} - \nonumber \\
& & -\partial_A \partial_B \phi- 
 e^{-\alpha}\left(\alpha^{\prime}-2 \sigma^{\prime}\right) \partial_A \partial_B B-e^{-\alpha} \partial_A \partial_B B^{\prime} + \nonumber\\ & & + \partial_A \partial_B \psi -\frac{1}{2} e^{-\alpha+\sigma}\left(\alpha^{\prime}-\sigma^{\prime}\right)\left(\partial_A B_B+\partial_B B_A\right)- \nonumber\\ & & -\frac{1}{2} e^{-\alpha+\sigma}\left(\partial_A B_B^{\prime}+\partial_B B_A^{\prime}\right) +\frac{1}{6} e^{6 \sigma} \partial_1^2\left(\partial_A \widetilde{B}_B+\partial_B \widetilde{B}_A\right).\nonumber
\end{eqnarray}
For the components of the perturbed stress-energy tensor, corresponding to radiation in the form of a perfect fluid paremeterized by quantities defined in (\ref{eq:mat11}), we have
\begin{eqnarray}
\delta T_{00}^{(\textrm{rad.})} &=& e^{2 \alpha} \overline{\rho}\left(\Delta+2 \phi\right),\\
\delta T_{01}^{(\textrm{rad.})} &=&
-e^\alpha \overline{\rho}\left[\partial_1 \chi+\frac{4}{3} e^{2 \alpha-4 \sigma} \delta u^1\right],\nonumber\\
\delta T_{0A}^{(\textrm{rad.})} &=&-e^\alpha \overline{\rho}\left[\partial_A B+e^\sigma B_A + \frac{4}{3}e^{2 \alpha+2 \sigma}\left(\partial_A\delta  u_{\parallel}+\delta u_{\perp }^A\right)\right],\nonumber\\
\delta T_{11}^{(\textrm{rad.})} &=&
\frac{1}{3}e^{2 \alpha-4 \sigma} \overline{\rho}(\Delta-2 \psi),\nonumber\\
\delta T_{1A}^{(\textrm{rad.})} &=&
\frac{1}{3} e^{2 \alpha+2 \sigma} \overline{\rho} \partial_1 \widetilde{B}_A,\nonumber\\
\delta T_{AB}^{(\textrm{rad.})} &=&
\frac{1}{3} e^{2 \alpha+2 \sigma} \overline{\rho} \Delta \delta_{A B}.\nonumber
    \end{eqnarray}
Conservation laws $T_{\mu\nu}^{(\textrm{rad.});\mu}u^{\nu}=0$, $T^{(\textrm{rad.});\mu}_{1\mu}=0$, and scalar and vector parts of $T^{(\textrm{rad.});\mu}_{A\mu}=0$ yield
\begin{eqnarray}
& & \Delta^{\prime} + \dfrac{4}{3}\left[-\psi^{\prime} + e^{\alpha}\left(\partial_1\delta u^1 + \partial_A\partial_A\delta u_{\parallel}\right)\right] = 0, \\
& & \partial_1\phi + \dfrac{1}{4} \partial_1\Delta + e^{-\alpha}\left(\partial_1\chi^{\prime}-\alpha^{\prime}\partial_1\chi\right) + e^{\alpha-4\sigma} \left[\delta u^{1\prime}+(\alpha^{\prime}-4\sigma^{\prime})\delta u^1\right] = 0, \nonumber\\
& & \phi + \dfrac{1}{4}\Delta - e^{-\alpha}\alpha^{\prime} B + e^{-\alpha} B^{\prime} + e^{\alpha+2\sigma} \left[(\alpha^{\prime}+2\sigma^{\prime})\delta u_{\parallel} + \delta u_{\parallel}^{\prime}\right] = 0, \nonumber\\
& & e^{-\alpha+\sigma}\left[(-\alpha^{\prime}+\sigma^{\prime})B_A + B_A^{\prime}\right] + e^{\alpha+2\sigma} \left[(\alpha^{\prime}+2\sigma^{\prime})\delta u_{\perp} + \delta u_{\perp}^{\prime}\right] = 0 \nonumber.
\end{eqnarray}
Components of the stress-energy tensor of the anisotropic solid remnant described through quantities defined in (\ref{eq:mat12}) are
\begin{eqnarray}
    \delta T_{00}^{(\Phi)} &=& \lambda\left[2f^0 e^{2 \alpha_0}  \phi + 2 c^2f_\Theta^0  \left(\psi+\partial_1\delta x\right)+c^2 f_{\Omega}^0 \partial_A^2 A_\parallel\right],\\
    \delta T_{01}^{(\Phi)} &=& \lambda\left[-f^0 e^{\alpha_0} \partial_1\chi+2 c^2 f_\Theta^0 \delta x^{\prime}\right],\nonumber\\
    \delta T_{0A}^{(\Phi)} &=& \lambda\left[-f^0 e^{\alpha_0}\left(\partial_A B+B_A\right)+c^2 f_{\Omega}^0 \left(\partial_A A_{\parallel}^\prime+ A_{\perp A}^\prime\right)\right],\nonumber\\
    \delta T_{11}^{(\Phi)} &=& 
\lambda\Big[ (2f^0 e^{2\alpha_0} - 2 c^2 f^0_{\Theta} + 4 c^4 f^0_{\Theta\Theta}e^{-2\alpha_0}) \psi + (2 c^2 f^0_{\Theta} + 4 c^4 f^0_{\Theta\Theta}e^{-2\alpha_0}) \partial_1 \delta x + \nonumber\\ & & + (c^2 f^0_{\Omega} + 2 c^4 f^0_{\Theta\Omega}e^{-2\alpha_0}) \partial_B\partial_B A_{\parallel}  \Big],\nonumber\\
\delta T_{1A}^{(\Phi)} &=& \lambda\left[-f^0 e^{2 \alpha_0} \partial_1 \widetilde{B}_A+2 c^2 f_{\Theta}^0 \partial_A \delta x + c^2 f_{\Omega}^0\left(\partial_1\partial_A A_{\parallel}+\partial_1 A_{\perp A }\right)\right],\nonumber\\
\delta T_{AB}^{(\Phi)} &=& \lambda  \Big\{\Big[\left(- 2 c^2 f_\Theta^0 + 2 c^4 f_{\Theta \Omega}^0e^{-2 \alpha_0} \right)\left(\psi + \partial_1 \delta x\right) + \nonumber\\ & & + \left(- c^2 f_{\Omega}^0 + c^4 f_{\Omega \Omega}^0 e^{-2 \alpha_0} \right) \partial_A^2 A_{\parallel}\Big] \delta_{A B} +\nonumber\\
& & + c^2 f_{\Omega}^0 \left(2 \partial_A \partial_B A_\parallel+\partial_B A_{\perp A}+\partial_A A_{\perp B}\right)\Big\}.\nonumber
\end{eqnarray}

{\setstretch{1.0}

}

\end{document}